\documentclass[apj,numberedappendix]{emulateapj}
\usepackage[english]{babel}
\usepackage{graphicx}
\usepackage{latexsym}
\usepackage{amsmath,amssymb}
\usepackage{dcolumn}
\usepackage{bm}
\usepackage{calrsfs}
\usepackage{ulem}
\usepackage[usenames]{color}
\usepackage{xspace}

\begin{document}

\newcommand{\na}{New Astr.}
\newcommand{\nar}{New Astr. Rev.}
\newcommand{\jcap}{JCAP}

\def\be{\begin{equation}}
\def\ee{\end{equation}}
\def\ba{\begin{eqnarray}}
\def\ea{\end{eqnarray}}
\def\d{\delta}
\def\e{\epsilon}
\def\f{\varphi}
\def\k{\varkappa}
\def\tde{\tilde}
\def\p{\partial}
\def\ms{\mathstrut}
\def\s{\strut}
\def\ds{\displaystyle}
\def\ts{\textstyle}
\def\b{\boldsymbol}
\def\r{\mathrm}
\def\G{\Gamma}
\def\sun{\odot}
\def\c454{3C454.3\xspace}
\def\ag{{\it AGILE}\xspace}
\def\fer{{\it Fermi}/LAT\xspace}
\def\pks{\object{PKS~2155$-$304}}
\def\opks{\object{PKS~1510$-$089}}
\def\mkn501{\object{Mkn~501}}

\title{Star-Jet Interactions and Gamma-Ray Outbursts from  3c454.3}

\author{D.V.~Khangulyan$^{1}$, 
M.V.~Barkov$^{2,3}$, 
V.~Bosch-Ramon$^{4}$,
F.A.~Aharonian$^{5,2}$,
\and 
A.V.~Dorodnitsyn$^{6,7}$
}

\affil{
$^{1}$Institute of Space and Astronautical Science/JAXA, 
3-1-1 Yoshinodai, Chuo-ku, Sagamihara, Kanagawa 252-5210, Japan,\\
$^{2}$Max-Planck-Institut f\"ur Kernphysik,
Saupfercheckweg 1, D-69117 Heidelberg, Germany\\
$^{3}$ Space Research Institute RAS, 84/32 Profsoyuznaya Street, Moscow, 117997, Russia\\
$^{4}$ Departament d'Astronomia i Meteorologia, Institut de Ci\`encies del Cosmos (ICC), Universitat de Barcelona (IEEC-UB),
Mart\'i i Franqu\`es, 1, E08028, Barcelona, Spain\\
$^{5}$Dublin Institute for Advanced Studies, 31
Fitzwilliam Place, Dublin 2, Ireland\\
$^{6}$Laboratory for High Energy Astrophysics, NASA Goddard Space Flight Center, Code 662, Greenbelt, MD, 20771, USA\\
$^{7}$Department of Astronomy/CRESST, University of Maryland, College Park, MD 20742, USA
}

\begin{abstract}
{We propose a model to explain the ultra-bright GeV gamma-ray flares observed from  the blazar \c454. The model is based on the
concept of a relativistic jet interacting with  compact gas  condensations produced when a star (red giant)  crosses the jet close to
the central black hole.  The study   includes an analytical treatment of the evolution of the envelop lost by the star
within the jet, and calculations of the related  high-energy radiation.
The model readily explains the day-long, variable on timescales of hours, GeV gamma-ray flare from \c454,
observed during November 2010 on top of a weeks-long plateau. In the proposed scenario,  the plateau state 
is caused by a strong wind generated by the heating of the star atmosphere by nonthermal particles accelerated at the jet-star interaction
region. The flare itself could be  produced by a few clouds of matter lost by the red giant after the initial impact
of the jet. In the framework of the proposed scenario, the observations   constrain the  key model parameters of the source, including  the 
mass of the central black hole: $M_{\rm BH}\simeq 10^9 M_{\odot}$, the total jet power: $L_{\rm j}\simeq 10^{48}\,\rm erg\,s^{-1}$,
and the Doppler factor of  the gamma-ray emitting clouds, $\delta\simeq 20$. Whereas we do not specify the particle acceleration mechanisms, 
the potential  gamma-ray production processes  are discussed and compared in the context of the proposed model. We argue that
synchrotron radiation of protons has certain advantages compared to other radiation channels of directly accelerated
electrons.}
\end{abstract}




\maketitle

\section{Introduction}
\label{intro}

\c454 is a powerful flat-spectrum radio quasar located at a redshift $z_{\rm rs}=0.859$. This source is 
very bright  in the GeV
energy range; during strong flares, its  apparent (isotropic) luminosity can reach $L_{\gamma}\gtrsim 10^{50}$~erg~s$^{-1}$
\citep[e.g.][]{agile10,fermi10,agile11,fermi11_3C}. The mass of the central black hole (BH) in
 \c454 is estimated in the range $M_{\rm BH}\approx (0.5-4)\times 10^9 \,M_{\odot}$ \citep{gcj01,bgf11}. This implies an Eddington luminosity
$L_{\rm Edd}\approx (0.6-5)\times 10^{47}\mbox{erg s}^{-1}$, which is several orders of magnitude below $L_\gamma$.
Although the  large  gap  within  $L_{\rm Edd}$ and $L_\gamma$  is naturally explained by  
relativistic Doppler  boosting,  the estimates of the jet power during these flares appear,
in any realistic scenario, close to or  even larger than  the  
Eddington luminosity \citep[][]{bgf11}.  Being  quite extreme, "super-Eddindton"  jets 
cannot be nevertheless  excluded for  accreting black holes  with very 
high accretion rates but low radiation efficiencies. 
Although \c454 is an exceptional case \citep[e.g. discussion in][]{bgf11},  data from other
objects provide additional evidence in favor of jets with super-Eddington mechanical power \citep{lp12}.  

The   GeV emission from \c454 is highly erratic, with variability  timescales as short as 3~hr, 
as reported, in particular,  for
the  December 2009 flare  \citep{fermi10}. The most  spectacular  flare regarding both 
variability and gamma-ray
luminosity was  observed in November 2010 by  \ag and \fer \citep{agile11,fermi11_3C} telescopes. 
During this high state, with the most active
phase lasting for 5 days, the apparent  luminosity in GeV  achieved  $L_{\gamma}\approx 2\times10^{50} \mbox{
erg s}^{-1}$. Around the flare maximum, the rising time was $t_{\rm
r}\approx 4.5$~hr, and the decay time, $t_{\rm f}\approx 15$~hr. The detection of photons with energies up to $\approx
30$~GeV, the short variability, and the contemporaneous X-ray flux constrain the Doppler boosting of the emitter to
$\delta_{\rm min}\gtrsim 16$ to avoid severe internal $\gamma\gamma$ absorption in the
X-ray radiation field \citep{fermi11_3C}.

A remarkable feature of the gamma-ray emission from \c454 is the so-called plateau phase 
revealed  during the bright flare  in 2010.    It is characterized by a long-term 
brightening of the source,   a  few weeks before the appearance of the main flare. Such  plateau states
have been  observed by \fer for  three flares
\citep[e.g.][]{fermi10,fermi11_3C}, with the plateau emission being about an order of magnitude  
fainter than that of the main flare.

Remarkably,  the rapid gamma-ray variability of \c454  is accompanied by an activity at lower energies. The simultaneous 
multiwavelength observations  of the source during flares  have revealed a strong correlation with optical and X-rays. It has been
interpreted as evidence that the gamma-ray source is located upstream from the core of the 43~GHz radio source, 
which is at a distance $z<$~few~pc from the central BH \citep[see, e.g.,][]{jml10,jor12,wmj12}. 

Over the recent years,  several  works have attempted to explain the flaring gamma-ray activity of \c454 
within the framework of the standard synchrotron self-Compton (SSC)
or external inverse-Compton (EIC) models \citep{kg07,gft07,smm08,bgf11}. 
In the SSC scenario,  it is possible to  reproduce
the spectral energy distribution (SED) from optical wavelengths to  gamma-rays. 
In these models  most of the jet power is (unavoidably) carried by protons,
and only a small fraction is contained  in relativistic electrons and the magnetic field.  
The required proton-to-Poynting flux
ratio of $L_{\rm p}/L_{\rm B}\sim 100$ is quite large. Such a  
configuration would be hard to reconcile, at least in the gamma-ray emitting region 
close to the central BH, with an undisturbed jet which is launched by the Blandford-Znajek  \citep{BZ77} type
(BZ) process, 
in which the luminosity of the jet is dominated by Poynting flux and the jet
consists of $e^\pm$-pairs  \citep[see also][]{ruffini75,lovelace76}. 
In this regard we should mention that recent relativistic magnetohydrodynamical (MHD) simulations of 
jet acceleration yield much less efficiency of  conversion of the  
magnetic energy into bulk motion kinetic  energy; these calculations  
\citep[][]{kbvk07,kvkb09,tnm10} predict a quite modest ratio  $(L_{\rm p}+L_{e^\pm})/L_{\rm B}\lesssim 4$.

As stars and clumpy matter are expected to be present in the jet surroundings,
they could be behind the powerful gamma-ray events in
AGN  \citep[see, e.g., a  discussion
in][]{bpb12}.  In particular, a red giant (RG) can enter into the jet,  lose its external layers, and in this way 
generate a strong perturbation inside the jet. This perturbation can accelerate particles and produce high-energy radiation. 

The  jet-RG interaction (JRGI) scenario has been invoked 
to explain the day-scale flares in the  nearby
non-blazar type AGN  M87 \citep{bab10},  
It  has been  applied also to the TeV blazar \pks \  \citep{babkk10} to demonstrate that 
the jet-driven acceleration of debris from the  RG atmosphere can explain ultra-fast variability of 
very high-energy gamma-ray emission  on  timescales as short as $\tau\sim 200$~s. 
A  distinct   feature of the JRGI scenario is  the  high magnetization ($L_{\rm B}/L_{\rm p,e}\gg  1$) of the relativistic 
flows located at sub-parsec distances, where the gamma-ray production 
supposedly takes place.  Although  the  strong  magnetic field, $B \geq 10$~G,  dramatically reduces the efficiency of  the inverse 
Compton scattering of electrons, it opens an alternative channel of gamma-ray production through synchrotron radiation of protons
\citep{ah00,mp01}. 
The latter can be effectively realized only in the case of acceleration of protons to the highest possible energies, 
up to $10^{20}$eV.  
Thus the second (somewhat ``hidden")   requirement of this model is a very effective acceleration of protons with a rate close to the
theoretical limit dictated by the classical electrodynamics \citep{abdkk02}.  Although somewhat speculative, this condition does
agree with  rather
model-independent  (derived from first principles) arguments  that  the relativistic jets in AGN present the best candidate sites of production of 
the highest energy cosmic rays \citep{abdkk02,lw09}.  The large magnetic field and acceleration of
protons to the highest possible energies, coupled with strong Doppler boosting in relativistic outflows, not only provide an extension of
the gamma-ray spectra up to TeV energies, but also provide variability as short as 1~h \citep{ah00}. 

It is interesting to note that also inverse Compton  models can be 
accommodated, at least in principle, in the JRGI  scenario. Moreover, unlike most of the leptonic models 
of powerful blazars, in which the  requirement of a very low magnetic field, 
implying a deviation from the equipartition condition by orders of magnitude, generally 
is not addressed and explained, the JRGI scenario  can offer
a natural way for leptonic models to be effective assuming that  the gamma-ray  emission 
is produced through the inverse Compton scattering in shocked clouds originally weakly   
magnetized \citep[see ][]{bba12}. 

In this work, we show that  the JRGI scenario gives a viable mechanism for the explanation of the flares
seen in \c454. 
We also  argue  that  within this model   the plateau state can form due
to the interaction of the jet with a stellar wind excited by 
nonthermal (accelerated)  particles that penetrate into the red giant atmosphere.
Given its extreme nature and importance, we will  use  the very powerful GeV flare of  \c454  occurred in 
November 2010 as a template for our interpretation. We will show that this  active period of \c454,
consisting  of a day-long flares with variability as short as a few hours  on top of a weeks-long plateau state, can be explained by the
interaction of a RG with the jet relatively close to the central black hole. 

\begin{figure}
\includegraphics[width=0.47\textwidth,angle=0]{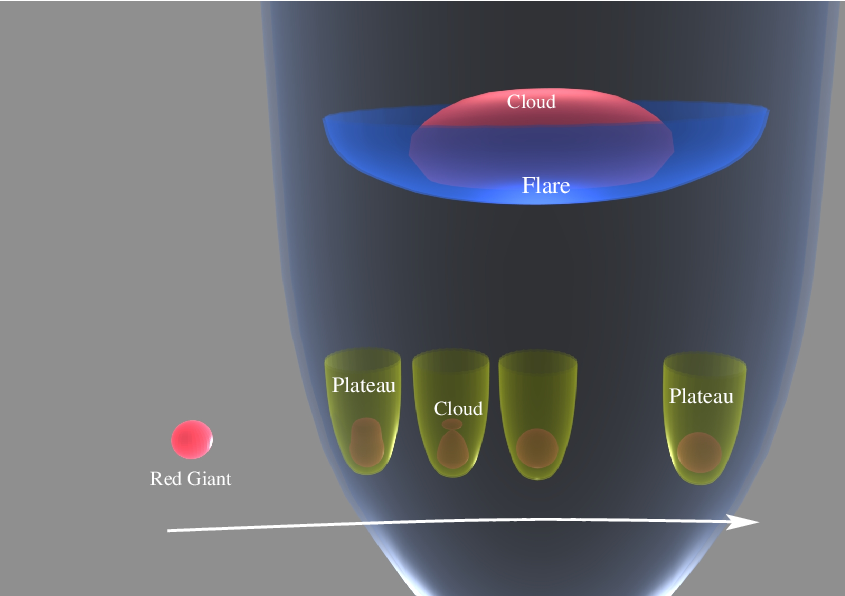}
\caption{Sketch for the JRGI scenario, in which a star moving from left to right penetrates into the jet. 
The star external layers are shocked and carried away, and a cometary tail, origin of the plateu emission, forms. 
The acceleration and expansion of the bigger clouds from the initially blown-up external layers of the star would 
lead to the main flare.}
\label{sketch}
\end{figure}

\section{Star-Jet interaction Scenario}
\label{sec:scenario}

The possibility that some exceptional flares in AGN may originate from star-jet 
 interactions has been proposed and discussed in our recent papers  
 \citep{bab10,babkk10,bba12}.  In this work  we explore the question 
 whether the extremelly  bright and short  gamma-ray flares  
detected   by the {\it Fermi} Large Area Telescope ({\it Fermi}/LAT)  from  \c454
in November 2010  can be explained by this model.  
It is important to note  that these flares  provide us  information of 
exceptional quality concerning  both the temporal behavior of the source and its energy
spectrum.  In this regard we should note that the operation of {\it Fermi}/LAT in scanning mode allows continuous monitoring of 
the source, so its temporal behavior can be studied without interruptions, including the 
pre- and post-flare epochs. 
Before considering the case of \c454, in this section we  discuss different
implications of the JRGI scenario regarding the general temporal structure of the active phase in a  blazar-type AGN.

When a star enters into an AGN jet, some stellar material is expected to be released inside the jet flow, eventually
forming a population of blobs that will be accelerated by the jet ram pressure. During this process, depending on 
the local  magnetic fields, different types of shocks and plasma waves  can be produced and 
propagate through both the jet material and the blobs. These waves
may accelerate particles to relativistic energies  the interaction of which with the ambient 
radiation and magnetic fields  would result in a  broadband nonthermal emission. Generally the properties 
of this emission  depend on the specifics of  the acceleration and radiation mechanisms and the details of the
target fields. However,   in the case of  extremely  effective particle acceleration and radiation, 
i.e. when the corresponding acceleration and radiative cooling times are shorter than other
timescales characterizing  the dynamics of the system (a mandatory condition 
given the enormous luminosity of the emission), the lightcurve of the emission will be essentially determined by the  jet/blob
interaction model. 

In the {\it fast cooling regime}, the proper intensity of  the nonthermal emission, i.e. the intensity 
in the blob co-moving reference frame, is  proportional to the energy released at the jet-blob interface. This energy release can be 
characterized by a simple dynamical model, which describes the acceleration of the blob by the jet ram pressure. In this model
there are just a few relevant parameters that describe the basic properties of the jet and the blob: the jet ram pressure ($P_{\rm j}$) and 
bulk Lorentz factor ($\Gamma_{\rm j}$), and the blob mass ($M_{\rm b}$) and  radius ($r_{\rm b}$; or, equivalently, its cross-section: $S_{\rm b}=\pi r_{\rm b}^2$) 
\citep[for details, see][]{babkk10}. 
The time dependence of the intensity of the jet/blob interaction
corrected for the Doppler boosting can be treated as a first-order approximation for the radiation
lightcurve. 

There are several important stages in the JRGI scenario that  may have an important impact on the observed lightcurve of radiation: 
(i) first, the removal of the  star external layers; (ii) the formation of a cloud from the removed stellar
material, which under the jet impact suffers heating and expansion, and eventually  fragments into a set of smaller blobs\footnote{Throughout the paper we will use the subscrips ``$c$'', ``$b$'', and ``$0$'' to  characterize the parameters of the initially formed cloud, of the blob, and of the jet at the star crossing height, respectively.};  (iii) the acceleration of  the
blobs in the direction of the jet flow  to a Lorentz factor equal or smaller than the jet Lorentz factor.  The intensity of 
the nonthermal processes associated with the blob motion has a strong time
dependence. At the start of the acceleration of the blob, its Lorentz factor is  modest, yielding a small (but growing with time)  
boosting of the flux. At later stages, when the blob Lorentz factor approaches the Lorenz factor of the jet, the radiation  intensity 
decreases  because of the weakening of the jet-blob interaction. According to the results obtained in  \citet{babkk10} in the fast cooling
limit, the maximum of the apparent luminosity, accounting for the Doppler boosting of the emission, occurs when the blob bulk Lorentz
factor achieves $\Gamma_{\rm b}\sim 0.8\Gamma_{\rm j}$.

During the penetration of the star into the jet, strong shocks are induced in the stellar atmosphere and a significant
part of the envelop can be removed. Although the details of this process might be very complicated and depend on the
radial  density profile of the star and the structure of the jet outer boundary, 
here, for simplicity, we assume an instantaneous penetration. 
To a certain extent, this simplification is justified by the stellar proper motion, which is expected to be faster than the initially 
induced  shock waves. Nevertheless, independent of the details of the star penetration,  one can expect that
when the star is fully within the jet,  a dense cloud will be released.  Later, as the cloud propagates through the jet, the
star may still release an intense wind or a sequence of small clouds due to ablation of the stellar atmosphere by the jet.

The mass of the cloud $\Delta M$ formed at the initial stage can be estimated by comparing the jet ram pressure, 
\begin{equation}
P_0\simeq \frac{L_{\rm j}}{c\pi\omega_0^2}\,
\label{eq:ram}
\end{equation}
with the gravitational force.  
Here, $\omega_0$ is the jet cross-section at the star crossing height. This gives  the
following estimate:
\begin{equation}
\Delta M=\frac{\pi P_0 R_{*}^4}{G M_{*}} 
\label{dms}
\end{equation}
where $M_{*}$ and $R_*$ are the RG mass and radius, respectively. 

The absolute upper limit on the continuous mass-loss rate
can be estimated comparing the energy flux density of the jet $q_0 = L_{\rm j}/\pi\omega_0^2$
and the gravitational force work density for removing matter from the stellar surface $q_{\rm hd}= \rho v G
M_{*}/R_{*}=\dot{M}G M_{*}/\left(\pi R_{*}^3\right)$. The maximum continuous mass ejection from the star is
\begin{equation}
\dot{M}= { c\pi P_0 R_*^3\over G M_*}.
\label{mdot}
\end{equation} 
A more precise estimate of this rate requires a detailed study of
the interaction process, which is out of the scope of this paper. Nevertheless, in
Sect.~\ref{pla} we discuss  whether a sufficiently strong stellar wind can be excited by the penetration of 
nonthermal (accelerated) particles.

Since the initial size of the expelled cloud should be comparable to the size of the star, it is possible to estimate the cloud
expansion time as $t_{\rm exp}\propto 2R_*/c_{\rm s}$, where $c_{\rm s}$ is the sound speed of the shocked material:
$c_{\rm s}\approx\left[(4\pi R_*^3/3)\gamma_{\rm g} P_0/M_{\rm c}\right]^{1/2}$. 
The cloud expansion time is 
\begin{equation}
t_{\rm exp} \approx A_{\rm exp} 
\left(\frac{M_{\rm c}}{\gamma_{\rm g} R_{*} P_0}\right)^{1/2},
\label{eq:expansion}
\end{equation}
where $\gamma_{\rm g}=4/3$ is the plasma adiabatic coefficient, and  $A_{\rm exp}$ is a constant of about a few 
\citep{gmr00,nmk06,phf10,bpb12}. According to the RHD simulation by \citet{bpb12}, a value of $1.5$ can be adopted for $A_{\rm 
exp}$.

The blob acceleration occurs on a timescale of \citep[see e.g. ][]{babkk10} 
\begin{equation}
t_{\rm acc}\approx\left\{{\frac{z_0}{c}\quad {\rm if}\quad D<1\atop\,\frac{z_0}{c}\frac{1}{D}\quad {\rm if}\quad D>1\,.}\right.
\label{dtpeak}
\end{equation}
The $D$--parameter that will be often used in the paper has a simple meaning. It is a dimensionless inverse mass of the blob: 
\begin{equation} 
D\equiv {P_{\rm 0}\pi r_{\rm b}^2z_0\over4c^2M_{\rm b}\Gamma_0^3}
\,. 
\label{DD} 
\end{equation} 
The above timescale corresponds to the blob acceleration in the laboratory reference frame. However, since the blob gets
accelerated towards the observer, the emission delay, as seen by the observer, should be approximately corrected by a 
factor of
$1/(2\Gamma_0^2)$. Thus, the observed peak of the emission should be delayed by a time interval of
\begin{equation}
\Delta t=t_{\rm exp}+t_{\rm acc}/(2\Gamma_0^2)\,.
\label{eq:delay}
\end{equation}
There are some uncertainties in this equation. In particular,  for
$D<1$, the blobs can travel a distance comparable to $z$ and the jet
properties at the dominant emission blob location may differ
significantly from the blob formation site. Thus, the detailed
evolution of the emitter can only be addressed properly through
numerical modeling.  However, despite the adopted simple approach,
some important conclusions can be derived. Namely, given the 
dependence of the two components at the RHS of equation~(\ref{eq:delay}) on 
the stellar material mass ($M_{\rm c}^{1/2}$ and $M_{\rm b}$,
respectively), the radiation  associated with the heavy cloud expelled
first will be delayed with respect to the emission produced by
lighter clouds formed later.

The emission produced by lighter clouds allows an estimation of the time
required for the star to cross the jet. Once the star enters into the jet, the process of jet-star interaction should
proceed steadily, with the production of these lighter clouds being roughly constant on average. 
Thus, the whole duration of the light cloud-associated 
emission, if observed, can be taken as a direct measurement of the jet crossing time $t_0\approx 2\omega_0/V_{\rm orb}$,
where $V_{\rm orb}\lesssim\sqrt{2GM_{\rm BH}/z_0}$ is the star velocity. This yields the following relation:
\begin{equation}
t_0\gtrsim \sqrt{2} {\omega_0z_0^{1/2}\over c r_{\rm g}^{1/2}}\,,
\label{eq:cross_time_independent}
\end{equation}
where $r_{\rm g}$ is the BH gravitational radius.
Adopting the paradigm of magnetically-accelerated jets (see Appendix, equations~(\ref{eq:ap_mdj}~--~\ref{eq:ap_mdj2})), it is 
possible
to derive a very simple expression for this timescale:
\begin{equation}
t_0\gtrsim 2^{3/2} z_0/c\,.
\label{eq:cross_time}
\end{equation}
In this way, the duration of the jet-star interaction is determined by the interaction distance from the central BH. 

The physical properties of the emitting blobs determine the available energetics and the maximum flux reachable 
in the considered scenario. In the blob comoving frame, the jet-star interface energy flux is defined by 
\begin{equation}
q_{\rm b}=\left(\frac{1}{\Gamma_{\rm b}^2}-\frac{\Gamma_{\rm b}^2}{\Gamma^4_{\rm j}}\right)\frac
{cP_{\rm j}}{4}\,.
\label{eq:proper_flux}
\end{equation}

Assuming a fixed efficiency $\xi$ in the blob reference frame
for the transfer of jet power to nonthermal gamma rays (where $\xi\ll 1$), and correcting for Doppler boosting, one can 
estimate the luminosity of a blob:
\begin{equation}
L_{\gamma}=4\xi c F_{\rm e} P_0 \Gamma_0^2 \pi r_{\rm b}^2\,,
\label{eq:lum}
\end{equation}
where the correction function $F_{\rm e}$ depends on time; or, equivalently, on the blob location in the jet (for a 
mathematical definition of this function, see Appendix, equation \ref{eq:ap_lum}). We note that the structure of the jet, i.e., 
the dependence of the jet Lorentz factor on $z$, determines the actual dependence of $F_{\rm e}$ on $z$ (see also Appendix for 
details).

The maximum value of $F_{\rm e}$ monotonically depends on the
$D$ parameter, approaching a value of $0.4$ if $D\gtrsim 1$ and being
$\sim 0.1$ for $D=0.1$. This relatively weak dependence allows us to derive the
maximum blob luminosity. Also it is possible to obtain
an  estimate of the total energy emitted by a blob 
or  an ensemble of sub-blobs as a result of the fragmentation of the original cloud ($M_{\rm c}=\sum M_{\rm b}$), 
\begin{equation}
E_{\gamma}\simeq 8\xi \bar{F_{\rm e}} M_{\rm b/c}c^2\Gamma_0^3\,,
\label{eq:toten}
\end{equation}
which accounts for the total energy transferred by the jet to a blob during the
acceleration process, $M_{\rm b}c^2\Gamma_0$, and for the anisotropy of
the emission due to relativistic effects represented by the  factor 
$\Gamma_0^2$ (see Appendix~\ref{ap:energy}, in particular equation \ref{eq:toten_ap}, for details). 
 

\section{The November 2010 Flare}

\subsection{General structure of the active phase}\label{3c}

The total apparent energy of the GeV gamma-ray radiation detected during the flare observed from \c454 in November 2010 was about
$E_{\rm tot}\approx L_{\gamma}\Delta t/(1+z_{\rm rs}) \approx 2.3\times 10^{55} \;\rm erg$. The exceptionally high flux during 
this period allows the derivation of a very detailed lightcurve, as seen from Figure~1 in \citet{fermi11_3C}. The nonthermal activity lasted
for $t_{\rm full}\sim 80$~days. The onset of the activity period was characterized by a plateau stage. During the first $t_{\rm pl}\sim
13$~days, a rather steady flux was detected, with an apparent luminosity $L_{\rm
pl}\approx 10^{49}$~erg~s$^{-1}$.
The plateau stage was followed by an exceptionally bright flare, the total duration which was $t_{\rm fl}\sim 5$~days, with a
rise time of $t_{\rm r}\sim 4.5$~h. The maximum flux reached was $7\times 10^{-5} \mbox{ ph cm}^2 \mbox{s}^{-1}$, which corresponds to a
luminosity of $L_\gamma\simeq 2\times 10^{50}\rm erg\,s^{-1}$. The final stage of the flare phase was characterized by variable emission
with a flux approximately a factor of $\sim 5$ weaker than the main flare, but still a factor of $\sim 2$ above the plateau level.


The observed luminosity of the plateau phase allow us to determine a lower limit on the star mass-loss rate, which can be
derived by differentiating equation~(\ref{eq:toten}):
\begin{equation}
\dot{M}_{*}\approx 10^{23} L_{\rm pl,49} \xi^{-1}\Gamma_{0,1.5}^{-3}\,\mbox{g s}^{-1}\,,
\label{mdotw}
\end{equation}
where $L_{\rm pl,49}=L_{\rm pl}/10^{49}\,{\rm erg~s}^{-1}$.
Comparing this requirement to equation~(\ref{mdot}), it is easy to
see that for typical parameters of RG stars, equation~(\ref{mdotw}) represents a modest 
(a few percent) of the maximum possible mass-loss rate.
%
%
In Sect.~\ref{pla} we will return to this issue.


To evaluate the feasibility of the JRGI scenario for the \c454 main flare, 
it is necessary to check whether  
the flux, the total energy release, and the flare delay with respect to the onset of the plateau,
are well described by equations~\eqref{eq:delay}, \eqref{eq:lum}, and \eqref{eq:toten} 
for a reasonable choice of jet/star properties. 
For example, equation~\eqref{eq:lum} can be rewritten as
\begin{equation}
P_0=8\times10^3 F_{\rm e,max}^{-1}L_{\gamma,50}\Gamma_{0,1.5}^{-2}S_{\rm b,32}^{-1} \xi^{-1}\rm \,erg\,cm^{-3}\,,
\label{eq:super_edd}
\end{equation}
where $L_{\gamma,50}$ is the 
peak luminosity of the flare normalized to $10^{50}\rm erg\,s^{-1}$, and $S_{\rm b,32}=\pi r_{\rm b}^2/10^{32}\rm cm^2$ the blob 
cross-section. 

A total energy budget of the flare of $\sim 2\times10^{55}\rm erg$ is feasible, according to
equation~(\ref{eq:toten}), if
\begin{equation}
M_{\rm c,30}\Gamma_{0,1.5}^3\approx {0.04E_{\rm \gamma,55}\over \xi\bar{F}_{\rm e}}\approx {0.1\over \xi\bar{F}_{\rm
e}}\,,
\label{mass_condtion}
\end{equation}
where $M_{\rm c,30}=M_{\rm c}/10^{30}$~g is the mass of the blown up RG envelop (i.e. the initially formed cloud). 
This requirement appears to be very close to the one provided by equation~\eqref{dms}, where the jet ram pressure is now 
substituted using equation~\eqref{eq:super_edd}:
\begin{equation}
M_{\rm c,max}\approx {5\times10^{29}\over F_{\rm e,max}}R_{*,2}^4 M_{*,0}^{-1}L_{\gamma,50}\Gamma_{0,1.5}^{-2}S_{\rm b,32}^{-1}\, \rm
g\,,
\end{equation}
where $R_{*,2}=R_*/10^2R_\sun$ and $M_{*,0}=M_*/M_\sun$, respectively.

The second term in equation~\eqref{eq:delay} is expected to be short compared to the
duration of the plateau phase, even for $D\sim0.1$, and thus the duration of the 
initial plateau phase constrains the expansion time (see equation~(\ref{eq:expansion})):
$$
t_{\rm exp}\approx5.4\times10^6 F_{\rm e,max}^{1/2} \xi^{1/2}\times
$$
\begin{equation}
M_{\rm c,30}^{1/2} R_{*,2}^{-1/2} L_{\gamma,50}^{-1/2}\Gamma_{0,1.5}S_{\rm b,32}^{1/2} \rm \, s\,.
\label{eq:ex_time_2}
\end{equation}


\citet{fermi11_3C} found that the emission of the main flare consisted of 5 components (see Figure~2 in
that work): a nearly steady contribution, like a smooth continuation of the plateau emission, and 4 sub-flares of similar duration
and energetics. Although the uniqueness of such a fit is not statistically assessed, it looks a good empirical description of
the data. In the framework of the JRGI scenario, such a description is very natural. 
The steady component would be
attributed to light clouds, continuously ejected by the star, and the four sub-flares would correspond to much heavier blobs
formed out of the blown-up stellar envelop during the initial stage.  On the other hand, the decomposition of the main flare
in four sub-flares implies a strict limitation on the variability timescale. The flare rise/decay timescales should be longer
than the blob light crossing time corrected for the Doppler boosting. 
Since the shortest variability scale was $\sim 5{\rm h}/(1+z_{\rm rs})\sim10^4$~s, 
the maximum possible size of the emitting blobs can be estimated as:
\begin{equation} 
r_{\rm b}\approx 10^{16}\Gamma_{\rm 0,1.5}\rm\,cm\,. \label{eq:cloud_size_del} 
\end{equation} 

If the jet is magnetically driven, this size constraint can be expressed through the mass of the central BH (see 
equation \ref{eq:ap_mdj}):
\begin{equation} 
{r_{\rm b}\over\omega} < 0.5\,M_{\rm BH,9}^{-1}\,, \label{eq:cloud_size} 
\end{equation} 
which is restrictive only in the case of $M_{\rm BH,9}\gg 1$. For $M_{\rm BH,9}\lesssim 1$, the blobs can cover the entire jet 
without violating the causality constraint.

In summary, if the flare detected with \fer was produced by an RG entering into the jet, the jet properties should satisfy  
equations~\eqref{eq:super_edd}, \eqref{mass_condtion} and \eqref{eq:ex_time_2}, which correspond to the restrictions imposed by
the flux level, total energy release, and the duration of the plateau stage, respectively. Interestingly, this set of
equations allows the derivation of a {\it unique} solution, which can constrain all the key parameters through the value of
the $D$ parameter:

\begin{equation}
P_0=3\times10^6\frac{F_{\rm e,max}^{1.5}D^{1.5}}{\bar{F}_{\rm e}^{2.5}\xi z_{0,17}^{1.5}}\,\rm erg\,cm^{-3}\,,
\label{eq:lum_final}
\end{equation}

\begin{equation}
M_{\rm c}=4M_{\rm b}=5\times10^{30}\frac{F_{\rm e,max}^{1.5}{D}^{1.5}}{\bar{F}_{\rm e}^{2.5}\xi z_{0,17}^{1.5}}\,{\rm g}\,,
\label{eq:mass_final}
\end{equation}

\begin{equation}
\Gamma_0=8\left( \frac{\bar{F}_{\rm e}z_{0,17}}{F_{\rm e,max}D}\right)^{0.5}\,,
\label{eq:gamma_final}
\end{equation}
and
\begin{equation}
S_{\rm b}=8\times10^{30}\frac{z_{0,17}^{0.5}\bar{F}_{\rm e}^{1.5}}{F_{\rm e,max}^{1.5}D^{0.5}}\,{\rm cm}^2.
\label{eq:size_final}
\end{equation}
The value of the ram pressure determined by equation (\ref{eq:lum_final}) can be compared to the value achievable in a 
magnetically-driven jet:
\begin{equation}
P_{\rm mdj}=2\times10^4 z_{0,17}^{-1} \frac{L_{\rm j}}{L_{\rm Edd}} \rm \,erg\,cm^{-3}\,,
\label{eq:mdj_rampressure}
\end{equation}
where $z$ is the distance from the central BH. It is seen from the comparison of \eqref{eq:lum_final} and \eqref{eq:mdj_rampressure}
that the solution found implies a jet with luminosity exceeding by a 
factor of $>10$ the Eddington limit. This requirement is also consistent with the lower limit on the jet luminosity
\begin{equation}
L_{\rm j}>cS_{\rm b}P_0=8\times10^{47}z_{0,17}^{-1}\xi^{-1}{D\over \bar{F}_{\rm e}}\rm erg\,s^{-1}\,,
\label{eq:lum_lower_limit}
\end{equation}
which exceeds the Eddington limit for the mass of the central BH $M_{\rm BH}\sim 5\times10^8 M_{\odot}$ \citep[see][and references therein]{bgf11}.
To assess the feasibility of such a strongly super-Eddington jet remains out of the scope of this paper, although we note that 
\citet{lp12} have presented observational evidence indicating that such jets may not be uncommon. We also note that the 
requirement of a high luminosity is less severe in the case of small $D$-values and that in this limit one can derive a constraint on
the central black hole mass combining equations \eqref{eq:cross_time_independent} and \eqref{eq:size_final}:
\begin{equation}
M_{\rm BH}>7\times10^8M_\sun z_{0,17}^{3/2}\,.
\end{equation}


The coherent picture emanating from the jet properties derived above
suggests that the JRGI scenario can be responsible for the flare
detected from \c454 for a solution of the problem with a reasonable set of model
parameters. This solution is designed to satisfy the
requirements for (i) the total energy; (ii) the peak luminosity; and (iii) the
duration of the plateau phase. Therefore, some additional observational
tests are required to prove the feasibility of the suggested
scenario. To address this issue, we discuss in Sect. \ref{sec:proc} the feasibility of different
radiation mechanisms to explain the observations under the inferred emitter conditions. Also, in Sect. 3.2 we explore whether the
stellar wind induced by the RG-jet interaction can be responsible for
the flux level detected during the plateau stage. Recall that, as already shown, the required
mass-loss rate is well within the limitations imposed by
equation~\eqref{mdot}. 

Finally, the flare raise time, which is related to the blob acceleration timescale 
(see equation~\ref{dtpeak}), can be calculated for the obtained solution. Interestingly, in the limit of small $D$-values, this 
timescale appears to be independent on $D$, the only remaining free parameter, and matches closely the detected raise time of 
$t_{\rm r}\sim4.5\rm h$:
\begin{equation}
t_{\rm acc}/\left(2\Gamma_{\rm b}^2\right)\simeq5{\rm h}\,.
\end{equation}
This agreement can be treated as a cross-check that shows the feasibility of the proposed scenario.

\subsection{The stellar wind as the origin of the  plateau emission}
\label{pla}


When the star penetrates into the jet, strong perturbations are generated in both the jet structure and the star external
layers. Strong shocks and other processes of energy dissipation take place in the impacted jet and stellar material,
generating conditions under which nonthermal particles can be accelerated. Here we will not specify the acceleration
mechanism, which depending on the acceleration region may be such as Fermi~I, stochastic acceleration, or magnetic reconnection. The
accelerated particles can either be advected away, while radiating, from the perturbed jet region, or penetrate into the
stellar atmosphere and thermalize heating the ambient plasma. The heating can induce a strong wind that will significantly
enhance the stellar mass-loss rate. A similar enhanced mass loss can occur as well through Compton-heated winds in high-mass
X-ray binaries and AGN due to heating by X-rays from an accretion disc \citep[e.g.][]{bs73,dkp08a,dkp08b}, although for the
mass-loss rate required to explain the plateau phase (see equation~\ref{mdotw}), the accretion disk X-ray flux seems to be too small.
On the other hand, the nonthermal particles produced at the star-jet interaction region can be enough to 
heat the stellar atmosphere. 
Since even nonthermal  particles  with low energy can effectively heat the atmosphere. The nonthermal emission may be undetectable
during the heating process. The heating energy flux can be estimated as
\begin{equation}
F_{\rm nt}=\chi c P_0=
3\times10^{16}\chi P_{0,6}\,\rm erg\,cm^{-2}\,s^{-1}\,,
\end{equation}
where $P_{0,6}={P_0/(10^6\rm erg\,cm^{-3})}$ is the normalized jet ram pressure, and $\chi=X_{\rm a}Y_{\rm dif}$ the nonthermal 
particle-to-jet power ratio within the stellar atmosphere, a
combination of the fraction of jet power channeled into nonthermal particles $X_{\rm a}$, and the fraction of the
nonthermal particles diffusing into the RG atmosphere $Y_{\rm dif}$. In the case of a jet interacting with a heavy
obstacle, the value of $X_{\rm a}$ could be potentially close to 1. 
The entrainment of particles into the RG atmosphere could be also high due to a strongly asymmetric diffusion process:
whereas the strong magnetic field in the jet may prevent particles from crossing back, the weaker magnetic field in the
other side would stimulate particles to diffuse deep into the stellar atmosphere. For all this, 
the effective value of $\chi$ may be of order 1.

The mass flux of a stellar wind excited by nonthermal particles of flux $F_{\rm nt}$ can be estimated as:
\begin{equation}\nonumber
\mu =  10^{-12} \, \alpha_{-12} \frac{ F_{\rm nt} R_{*}^{1/2}}{ (GM_{*})^{1/2}}=
\end{equation}
\begin{equation}
7\times10^{-3}\alpha_{-12}R_{*,2}^{1/2}M_{*,0}^{-1/2}\chi {P_{0,6}}\,\rm{g}\,\rm{s}^{-1}\,{\rm cm}^{-2}, 
\label{muw}
\end{equation}
where $\alpha$ is related to the properties of the plasma heating and cooling. In the case of X-ray heating, one finds 
that $\alpha\sim
0.03/c$ or $\alpha_{-12}\sim 1$ \citep[i.e.][]{bsh77}. However, for nonthermal-particle heating, the efficiency of
transferring jet energy into the stellar wind can be higher because these particles can penetrate deeper into the stellar
wind. Thus, in this case $\alpha_{-12}$ can significantly exceed 1.
The induced mass-loss rate therefore will  be $\dot{M}_{\rm w}= \pi \mu
R_*^2$, or 
\begin{equation}
\dot{M} \approx 
10^{24}\alpha_{-12}R_{*,2}^{5/2}M_{*,0}^{-1/2}\chi P_{0,6}\,
\mbox{g s}^{-1}\,.
 \label{mdotx}
\end{equation}
To explain the plateau emission by the interaction of this induced stellar wind with the jet, the mass flux from 
equation~(\ref{mdotx}) should be greater than the one from equation~(\ref{mdotw}).  Adopting the typical parameters of the November 2010
flare, this requirement can be satisfied if the
following condition holds (the function $\bar{F}_{\rm e}$ here characterizes the main flare episode):
\begin{equation}
\alpha_{-12}\chi \gtrsim 2\bar{F}_{\rm e} R_{*,2}^{-5/2}M_{0,*}^{1/2}.
 \label{rrg}
\end{equation}
Formally, for the adopted RG normalization parameters, the required heating efficiency
should be rather high $\chi\approx 0.7$ (accounting for
$\bar{F}_{\rm e}<0.3$, as found for conical jets --see Appendix
for details--). However, a significantly lower heating
efficiency can be sufficient to fulfill equation~(\ref{rrg}) under
certain circumstances; $\chi$ could be easily reduced if
$D\ll1$ ($\bar{F}_{\rm e}\approx D$ in this regime), or assuming
that nonthermal particles contribute strongly to the pressure of the stellar atmosphere
 ($\alpha_{-12}\gg 1$), or for a RG radius exceeding the fiducial value $R_*=100\,R_\sun$. 
For instance, to reduce $\chi$
to $0.1$, one may adopt either $R_*=200R_\sun$; or
$\alpha_{-12}\approx 5$ and $R_*=100R_\sun$; or
$D\approx 0.05$. A combination of all these factors could further
decrease the required heating efficiency to lower values.  Therefore,
an induced stellar wind seems to provide a feasible way to generate the plateau
emission component.

\section{Radiation Mechanisms}
\label{sec:proc}
\subsection{General Comments}\label{genan}
In this section we discuss the feasibility of different radiation mechanisms for the conditions in the production site inferred 
in the previous section (see equations \ref{eq:lum_final}~--~\ref{eq:size_final}). The efficiency of the radiation channels is 
determined by the presence of nonthermal particles with the required energy and targets. Several types of target fields are 
related to efficient high-energy processes: matter in the case of proton-proton or bremsstrahlung mechanisms; magnetic field 
for synchrotron radiation; and photons for inverse Compton and photo-meson emission. The obtained solution for the properties 
of the jet, plus observational constraints, allow two of these targets to be properly characterized.

Since the mass and size of the blobs are estimated, one can derive the matter density in the production regions:
\begin{equation}
n_{\rm b}\sim 10^{7}\,\rm cm^{-3}\,.
\label{eq:n_density}
\end{equation}
This density estimate allows one to discard the proton-proton channel, since the expected cooling time,  
$t_{\rm pp}\approx 10^{15}/n \mbox{ s }\sim 10^{9}$~s, is too long\footnote{We note however that this limitation does not apply to the case of ``off-axis'' AGNs, since in that case there are no constraints related to the requirement of blob acceleration. In particular, proton-proton interactions were shown to be a feasible channel for the interpretation of the TeV emission detected from M87  \citep{bab10,bba12}}. This conclusion is also
valid to exclude the bremsstrahlung 
channel, with a similar cooling time.

\begin{figure*}
\vspace{0.6cm}
\includegraphics[width=0.99\textwidth,angle=0]{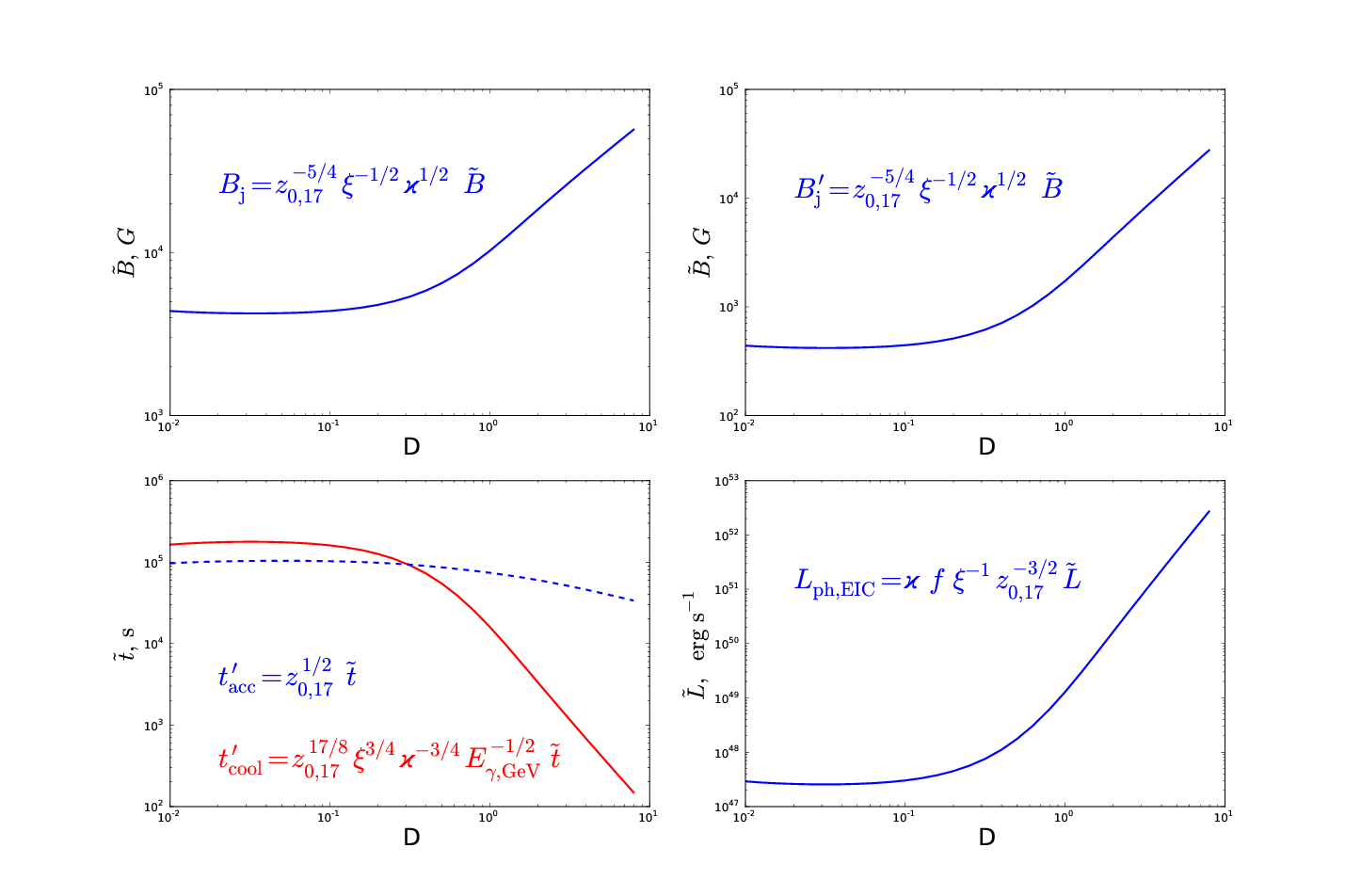}
\vspace{0.5cm}
\caption{Numerical calculations of the parameters related to radiation processes for the case of a conic jet vs the value of 
$D$. Top left panel: jet magnetic field; top right panel: jet magnetic field in the co-moving frame; bottom left panel: proton 
synchrotron cooling time (red solid line) and blob acceleration time (blue dashed line),  both in the co-moving frame; bottom 
right panel: the registered luminosity of the target photon field for the EIC scenario.}
\label{fig:rad_limits}
\end{figure*}

The derived ram pressure of the jet allows the derivation of an upper limit on the magnetic field strength:
\begin{equation}
B_{\rm j}<6\times10^3\frac{F_{\rm e,max}^{3/4}D^{3/4}}{\bar{F}_{\rm e}^{5/4}\xi^{1/2} z_{0,17}^{3/4}}\rm \,G\,.
\label{eq:b_field}
\end{equation}
In fact, from equipartition arguments, the actual strength of the magnetic field should be close to that value; i.e. one may 
take $\k\sim 1$ as the fiducial value for the fraction of the jet luminosity carried in electromagnetic form. Also, accounting for 
the relation ${F_{\rm e,max}^{3/4}D^{3/4}/\bar{F}_{\rm e}^{5/4}}\geq0.7$, valid in the range of feasible values of the 
$D$ parameter for the conical jet model, one can derive a lower limit on the jet magnetic field of
\begin{equation}
B_{\rm j}>3\times10^3\xi^{-1/2}\k^{1/2} z_{0,17}^{-3/4}\rm \,G\,.
\label{eq:b_field_jet}
\end{equation}
More detailed calculations of this value are presented in Fig.~\ref{fig:rad_limits}. In this figure we also show the value of 
the magnetic field in the comoving system ($B_{\rm j}'$), which was calculated using the jet Lorenz factor 
equation~\eqref{eq:gamma_final} as a function of the value of $D$. It can be seen that for the $D$-range of interest, there is a lower limit on
the magnetic field strength:
\begin{equation}
\begin{split}
B_{\rm j}'=B_{\rm j}\Gamma_{\rm b}^{-1}>&B_{\rm j}\Gamma_0^{-1}\\
>&200\xi^{-1/2}\k^{1/2}z_{0,17}^{-5/4}\rm \,G\,.
\label{eq:b_field_comove}
\end{split}
\end{equation}

Regarding the density of the target photon field, it cannot be constrained in general. However, in the specific case of the 
synchrotron-self Compton scenario, it is possible to put certain limitations. Namely, the high-energy component detected by \fer 
exceeds by two orders of magnitude ($f\approx100$) the flux detected at other wavelengths. This implies that the energy density 
of the target field should exceed by this factor the energy density of the magnetic field: $w_{\rm ph}'=fw_{\rm B}'$ (these are 
co-moving reference frame values). The luminosity of such a field can be estimated as
$$
L_{\rm target}\sim 4\pi r_{\rm b}^2 c w_{\rm ph}'\Gamma_{\rm b}^4\sim4\pi r_{\rm b}^2 c f w_{\rm B}'\Gamma_{\rm b}^4\sim
$$
\begin{equation}
4\pi r_{\rm b}^2 cf\k P_0\Gamma_{\rm b}^2\sim F_{\rm e}^{-1}\left({\Gamma_{\rm b}\over \Gamma_{0}}\right)^{2}\xi^{-1}f\k 
L_\gamma\,.
\end{equation}
Since $fL_{\rm target}<L_\gamma$ and $F_{\rm e}\simeq0.5\left({\Gamma_{\rm b}/\Gamma_{0}}\right)^{2}$, one obtains
\begin{equation}
\k f^2<0.5\xi \,.
\label{eq:ssh_constraint}
\end{equation}
Thus, the SSC scenario can be realized only in a specific region of the jet, where the magnetic field is significantly lower 
than the characteristic one (i.e., $\k<10^{-4}$). The blobs themselves may have a magnetic field quite different from the field 
in the jet, and therefore may serve as good sites for SSC. However, this assumption involves further
complexity, e.g. regarding internal shock acceleration within the blobs, and is therefore deferred to future studies. Also, we 
note that the radiation efficiency of photomeson production is low \citep[e.g. see discussion in][]{akc08}. If photomeson production 
were the radiation mechanism, it would imply a 
very low value of $\xi$ and an uncomfortably high power of the jet.

We focus now on the following radiation mechanisms: electron and proton synchrotron, and external inverse Compton. The 
analysis of the synchrotron channel is straightforward, since the density of the target is determined. Indeed, to produce a 
gamma ray of energy $E_{\rm \gamma,GeV}=E_{\rm \gamma}/1$~GeV it is required either an electron or proton of energy
\begin{equation}
E_{\rm e/p}=2\left(\frac{m_{e/p}}{m_{e}}\right)^{3/2}\xi^{1/4}\k^{-1/4}z_{0,17}^{3/8}E_{\rm \gamma,GeV}^{1/2}\rm \,TeV\,,
\end{equation}
We note that the weak D-dependence in these equations is neglected for simplicity.

To obtain such a high-energy particles a few conditions should be satisfied, the most critical ones being the {\it Hillas 
criterion} and the efficiency of the acceleration process.  The Hillas criterion requires the gyroradius of the highest energy 
particles to be smaller than the size of the acceleration region. Since the acceleration region size is determined by the blob 
cross-section, one can derive the following requirements:
\begin{equation}
\xi^{3/4}\k^{-3/4}z_{0,17}^{11/8}E_{\rm \gamma,GeV}^{1/2}<10^{8}\left({m_{\rm e}\over m_{\rm e/p}}\right)^{3/2} \,.
\end{equation}
This estimate allows one to illustrate that for the derived jet parameters the Hillas criterion appears to constrain the acceleration of particles in 
neither the lepton nor the proton synchrotron scenarios.

Regarding the efficiency of the
acceleration process, it is convenient to present the nonthermal particle acceleration time in a form independent of the 
specific acceleration
mechanism:
\begin{equation}
t_{\rm par}=\frac{\eta R_{\rm gy}}{c}= \frac{\eta E_{\rm e/p}}{eB_{\rm j}'c}\,
\label{acctime}
\end{equation}
where $E_{\rm e/p}$ is the energy of particle, $\eta\ge 1$ the acceleration efficiency, and $R_{\rm gy}$ the particle gyroradius. 
This timescale should be compared to the dominant cooling time (synchrotron) to determine the highest achievable energy:
\begin{equation}
t_{\rm sy}= \frac{9 m_{e/p}^4c^7}{4e^4 E_{e/p}B_{\rm j}'^2}>t_{\rm par}\,.
\label{cooltime}
\end{equation}
Accounting for the  energy of the emitted gamma-rays, $E_\gamma$, one can obtain an upper limit on the acceleration 
efficiency:
\begin{equation}
\eta< 5\times 10^3\Gamma_{\rm b,1} \left({E_{\gamma}\over m_{\rm e/p}c^2}\right)^{-1}\,,
\end{equation}
where $\Gamma_{\rm b,1}=\Gamma_{\rm b}/10$ is the blob Lorentz factor.
Since the red-shift corrected energy of the spectral break appears to
be close to $2$~GeV, the leptonic scenario can be realized only if
$\Gamma_{\rm b}>10\eta$. In the case of proton-synchrotron models this
constraint gets significantly relaxed yielding $\eta< 200\Gamma_{\rm
  b}$. However, given the relatively small values of the obtained
Lorentz factors (see equation \ref{eq:gamma_final}), the derived limits
on the acceleration efficiency are very tight, especially for the
leptonic case ($\eta\rightarrow1$).  The feasibility of such a scenario
should be addressed with a detailed acceleration model, which remains
out of the scope of the present paper. We note that this
limitation can be relaxed adopting a scenario with a highly
turbulent magnetic field \citep[see more details in ][]{kak13}.


Since the proton synchrotron scenario fulfills the basic requirements for the production mechanism, it is worthy to check 
whether one can reproduce the observed basic spectral properties. \fer has reported a broken powerlaw spectra with a nearly constant 
break energy. The change of the powerlaw index is varying being between 0.5 and 1. Whereas index changes of 0.5, thought to be 
cooling breaks, are typical in synchrotron scenarios, other values for the powerlaw index change may be hard to explain through cooling. On the 
other hand, one can attribute this spectral feature to absorption  \citep[as suggested for this source][]{sp11}. 
Regarding the uncooled part of the synchrotron spectrum, it is typically characterized by a power-law index of 1.5 
($n_\nu\propto \nu^{-1.5}$). Interestingly, the extrapolation of such a slope from the level found by {\it Swift} in X-rays 
gets close to the spectral points reported by the \fer Collaboration. Therefore, it is natural to expect that the cooling break 
is located at energies $\sim 10-100$~MeV. The synchrotron cooling time of protons in the comoving frame can be expressed as
\begin{equation}
t_{\rm cool}'\lesssim10^5\,E_{\rm \gamma,GeV}^{-1/2}\xi^{3/4}\k^{-3/4}z_{0,17}^{17/8}\rm s\,.
\end{equation}
This timescale should be compared to the timescale characterizing the comoving blob acceleration time:
\begin{equation}
t_{\rm acc}'\sim10^5 z_{0,17}^{1/2}\rm\,s\,.
\end{equation}
In figure \ref{fig:rad_limits} (bottom left panel) we show the
numerical computation of these timescales for the case of a conical
jet. One can see that for the typical model parameters, these time
scales are comparable. This allows us to obtain three important
conclusions for the considered scenario. Namely, that (i) the fast
cooling regime assumption should not be strongly violated in
proton-synchrotron models for GeV emitting protons; (ii) one should
expect a cooling break in the high-energy domain; and finally that (iii)
the proton synchrotron mechanism appears to be an efficient radiation
channel. Indeed, it is often argued based on the comparison of the proton
cooling and jet dynamical time scales \citep[see, e.g.,][]{s10psyn}
that proton synchrotron is characterized by a very low efficiency as
gamma-ray production mechanism in AGN jets. However, as it can be seen
from figure \ref{fig:rad_limits} (bottom left panel), for the inferred
jet properties, proton synchrotron can render a high radiation efficiency.


The external inverse Compton channel is not strongly constrained in the JRGI scenario, since the properties of the photon target are 
not well known and can vary within a broad range. Similarly to the SSC case, the density and luminosity of the target 
photons can be estimated based on the ratio of the synchrotron and IC luminosities. This yields the following luminosity
\begin{equation}
\begin{split}
L_{\rm ph, EIC}&\sim 4\pi z^2 c w_{\rm ph}\\
&>3\times10^{47}f\k \xi^{-1}z_{0,17}^{-1.5}\rm\, erg\,s^{-1}\,.
\end{split}
\end{equation}
The value of this luminosity as a function of $D$ is shown in figure \ref{fig:rad_limits} (bottom right panel). It is seen that 
the photon field luminosity required for the EIC scenario to work appears to be very high (accounting for 
$f\sim100$~\footnote{A similar consideration to the one done in the context of SSC applies here regarding EIC within the cloud, 
i.e. where potentially $\k\ll 1$.}), exceeding the reported luminosity of the BLR unless $\k$ is very small or the interaction 
region is located far enough from the central black hole ($z\gg10^{17}\rm \, cm$).

There is another constraint related to the process of gamma-ray absorption through pair creation in the local radiation 
fields. The peaking energy of the radiation component  detected with \fer should be close to 100~MeV. Otherwise, the lower 
energy spectrum should be very hard not to violated the X-ray flux detected with {\it Swift} \citep{fermi11_3C}. In the emitter 
comoving system this would correspond to
\begin{equation}
E_{\rm ph}'=10\,\Gamma_{\rm b,1}^{-1}\rm \, MeV\,,
\end{equation}
i.e. relatively close to the electron-positron creation threshold.  Assuming a photon spectrum $\propto E^{-1.5}$ below the peak (it may be
even harder; see Sect.~\ref{sedlc}), it is 
possible to compute the luminosity transferred to the secondary pairs:
\begin{equation}
L_{\rm sec}\sim5\times10^{48}L_{\gamma,50}^2\Gamma_{\rm b,1}^{-4}r_{\rm b,15}^{-1}\rm\, erg\,s^{-1}\,,
\label{eq:sec_lum}
\end{equation}
where $r_{\rm b,15}=r_{b}/10^{15}\rm \, cm$ is the normalized blob radius.
Substituting here the obtained solutions for the jet parameters (equations~\ref{eq:lum_final}-\ref{eq:size_final}) in the limit of small 
$D$, one gets
\begin{equation}
L_{\rm sec}\sim10^{49}z_{0,17}^{-2.25}\rm\, erg\,s^{-1}\,.
\label{eq:sec_lum2}
\end{equation}
This energy should be emitted either via synchrotron or inverse Compton radiation. 
The obtained estimate of the proper magnetic field from equation~\eqref{eq:b_field_comove} allows one to estimate the peaking 
energy for the synchrotron channel to be
\begin{equation}
\hbar \omega_{\rm sec,sy}\sim10^{-2}\xi^{-1/2}\k^{1/2}z_{0,17}^{-7/4}\rm \, eV\,.
\end{equation}
This component can be constrained with the infrared flux detected during the flare \citep{jor12}. If the dominant channel is 
inverse Compton scattering, the peak energy can be estimated as:
\begin{equation}
\hbar \omega_{\rm sec,ic}\sim4 \left({\epsilon_{\rm EIC}\over 40\rm \, eV}\right)\rm \, MeV\,,
\end{equation}
where $\epsilon_{\rm EIC}$ is the target photon energy. Although this component would peak above the X-ray energy band, the 
lower energy tail may give an important contribution to the reported flux level. Also, it is important to note that the inverse 
Compton photons may serve as targets for the absorption of primary gamma-rays leading to a non-linear regime of the 
emission formation. This, in particular, may make a self-consistent treatment of the EIC scenario very complex, and thus we 
leave this possibility for future dedicated studies. Nevertheless,
we wil account for internal gamma-ray absorption in the blob own synchrotron field, and comment on radiation 
reprocessing outside the blob in Sect.~\ref{sedlc}.

Despite being not very detailed, and order-of-magnitude, the above analysis shows that explaining the November 2010 flare 
detected from \c454 in the framework of the JRGI scenario requires a jet with a very high ram pressure. Consequently, a 
magnetic field with a strength not far from equipartition, say a factor of 10 below, appears to be too strong for radiation 
processes other than synchrotron. On the other hand, leptonic synchrotron emission requires an extremely efficient acceleration 
process with an acceleration parameter $\eta<2$, which cannot be discarded but is perhaps unrealistic. Thus, {\it proton 
synchrotron} emission seems the most comfortable radiation channel given the restrictions imposed by data and the JRGI 
scenario. In what follows we test this process using a simple radiation model.

\subsection{Modeling the Lightcurve and the Spectrum}
\label{sedlc}

To check whether JRGI plus synchrotron radiation can explain the
observations in the case of magnetically dominated jets (i.e., $\k=1$), we have computed the lightcurve of the November 2010
flare and the SED for one of its subflares.  The radiation output was
assumed to be dominated by proton synchrotron, being external or
synchrotron self-Compton neglected due to the strong magnetic field
(see Sect.~\ref{genan}).

To derive the lightcurve, equation~\ref{eq:lum} has been used (see the
determination of $F_e$ in the Appendix). In Figure~\ref{lc}, a
computed lightcurve that approximately mimics the November 2010 flare
is presented. The lightcurve has been obtained assuming four subflares
of total (apparent) energy of $10^{55}$~erg each, plus a plateau
component with luminosity of $2\times 10^{49}$~erg~s$^{-1}$. For
each subflare, we have adopted $D=0.1$. The normalization of
the lightcurve has been determined adopting the following values: the
Lorentz factor $\Gamma_0=28$, the ram pressure $P_{\rm j}=
3\times 10^6$~erg~cm$^{-3}$, blob radius $r_{\rm b}=2.7\times
10^{15}$~cm and $\xi=0.3$.  These parameters imply a minimum
jet luminosity of $L_{\rm j}=2.3\times 10^{48}$~erg~s$^{-1}$. The
remaining parameters for the emitter are $z_0=1.3\times 10^{17}$~cm
 and $M_{\rm b}=1.3\times10^{30}$~g. The
corresponding mass of the matter lost by the RG to explain the four
subflares is $5\times 10^{30}$~g, not far from the upper-limit given in
equation~\ref{dms}.

To calculate the SED, we have adopted a spectrum for the injected
protons $Q\propto E^{-p}\exp(-E/E_{\rm cut})$, and an homogeneous
(one-zone) emitter moving towards the observer with Lorentz factor
$\Gamma_{\rm b}=12$. The minimum proton Lorentz factor has been taken
equal to the shock Lorentz factor in the blob frame, i.e.  $E_{\rm
  min}=\Gamma_{\rm 0}/\Gamma_{\rm b} m_{\rm p}c^2$. The cutoff energy,
$E_{\rm cut}$, has been obtained fixing $\eta=4\times 10^3$
(see equation \eqref{acctime} for details), i.e., a modest
acceleration efficiency. For the maximum proton energy, i.e. how far
beyond the cutoff the proton energy is considered, we adopted two
values: $E_{\rm max}=\infty$ and $E_{\rm max}=3E_{\rm cut}$. Regarding
the latter case, we note that assuming a sharp
high-energy cut is very natural. The injection spectrum was selected
to be hard, $p=1$, to optimize the required energetics. Interestingly,
magnetic reconnection, in particular in magnetized jet-cloud
interactions, has been postulated to provide hard particle spectra
\citep[see, e.g.,][and references therein]{bos12}.

In Figure~\ref{sed}, the SED of a subflare is shown. The impact of the
internal absorption on the gamma-ray spectrum is negligible, although
the emission of the secondary pairs appears in the energy band
constrained by optical measurements \citep{jor12}. For the chosen
model parameters, the synchrotron secondary component goes right
through the optical observational constraints, and for slightly higher
$z_0$-values, the secondary emission will be well below the optical
points (see equation \eqref{eq:sec_lum2}).  Also, we note that the obtained spectrum does not violate the
X-ray upper-limits obtained by {\it Swift}.

To illustrate the impact of external $\gamma\gamma$ absorption, we
have introduced a photon field peaking at 40~eV with a luminosity
$4\times 10^{46}$~erg~s$^{-1}$, produced in a ring with radius
$10^{18}$~cm at $z=0$ around the jet base. Two photon fields have been
adopted, a black body and one represented by a $\delta$-function, to
simulate the impact of a dominant spectroscopic line.  As seen in
Figure~\ref{sed}, the impact of such an external field can be
important. The treatment of the secondary emission of the produced
pairs is beyond the scope of this work. However, we note that the
ambient magnetic field energy density could be well below that of
radiation, allowing electromagnetic cascades to develop, effectively
increasing the gamma-ray transparency. In fact, if it were not the case, the secondary
synchrotron emission would likely violate the constraints from Swift
data \citep[see, e.g.,][]{akc08,sit10,zck10}. EBL absorption is not
relevant in the energy range of interest.

As seen in figure 4, the spectral breaks induced by external
absorption are very sharp and do not fit the \fer data
points. However, the properties of the external photon field can vary
in a quite broad range, and it is expected that for some feasible
photon field one could achieve a very good agreement between the
emission spectrum and \fer measurements \citep[see,
e.g.,][]{sp11}. Instead of searching for such a field configuration,
here we consider an alternative possibility. As mentioned above, we
assume that the injection spectrum has not only an exponential cutoff
at $E_{\rm cut}$, but also a sharp upper limit at $E_{\rm max}=3E_{\rm
  cut}$.  As seen in figure 4, this assumption, natural accounting for
the fact that particles cannot have arbitrarily high energies in the
source, allows one to qualitatively model \fer observations without invoking
additional assumptions regarding external absorption.

In addition to optical photons, radio emission was also detected at
the flare epoch and thought to be linked to the gamma-ray activity
\citep{jor12}. This radiation is strongly sensitive to the details of
the flow dynamics, and at this stage we will not try to interpret
radio observations. However, we note that the energetics involved in
gamma-ray production is very large, and JRGI comprehends complex
magnetohydrodynamical and radiative processes, so it could easily
accommodate the presence of a population of radio-emitting electrons
at $z\ge z_{\rm flare}$.

{\it Swift} X-rays could be also linked to the JRGI activity. As stated in
Sect.~\ref{genan}, X-rays may come from secondary pairs produced via
pair creation, or from a primary population of
electrons(/positrons). However, as with radio data, given the
complexity of the problem we have not tried at this stage to explain
the X-ray emission contemporaneous to the GeV flare.


\begin{figure}
\includegraphics[width=0.43\textwidth,angle=0]{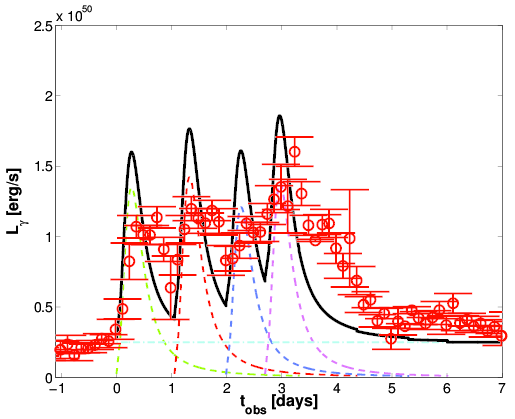}
\caption{Lightcurve computed adopting the parameters $L_{\rm j}=2.3\times 10^{48}$~erg~s$^{-1}$, $z=1.33\times 10^{17}$~cm 
$\Gamma_{\rm j}=28$, $M_{\rm c}=1.3\times10^{30}$~g, $r_{\rm c}=2.7\times 10^{15}$~cm, and $\xi=0.3$. 
We show 4 subflares  (dashed lines), plateau background
(dot-dashed line), and the sum of all of them (solid line). The observational data points and error bars are obtained from the 
\fer 3h binned count rates and photon index using luminosity distance of $D_{\rm L}=5.5\rm Gpc$ and assuming a pure powerlaw 
spectrum between 0.1 and 5~GeV.}
\label{lc}
\end{figure}

\begin{figure}
\vspace{0.6cm}
\includegraphics[width=0.43\textwidth,angle=0]{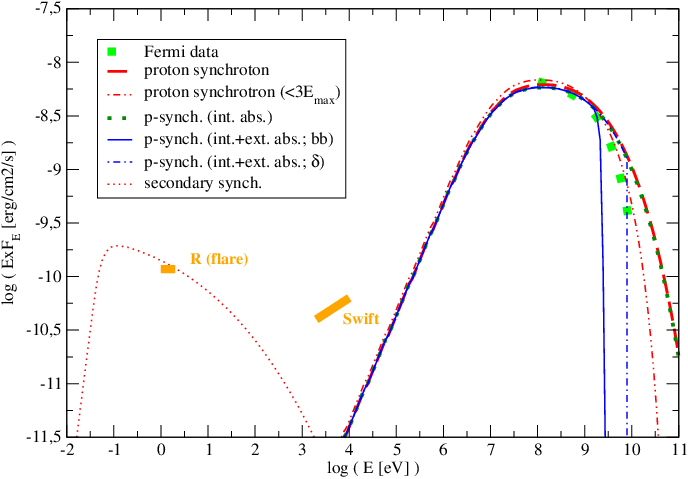}
\vspace{0.5cm}
\caption{Computed SED of the synchrotron emission for a subflare of the 2010 November. The thick dashed line shows intrinsic gamma-ray emission for the case of $E_{\rm max}=\infty$. Dotted and dot-dot-dashed line shows gamma-ray spectra corrected for internal absorption only for $E_{\rm max}=\infty$ and $E_{\rm max}=3E_{\rm cut}$, respectively.  The
thin solid and the dot-dashed lines correspond to the cases when absorption is dominated by a black body and a monoenergetic 
photon field, respectively.
The computed synchrotron SED of the secondary pairs produced via internal pair creation is also shown (dotted line).
The parameters of the flare are the same as in Figure~\ref{lc}. 
The shown observational data are from \fer, Swift 
\citep{fermi11_3C}, and the flux in the R band \citep{jor12}.}
\label{sed}
\end{figure}

\section{Discussion}

Observations in the VHE domain put severe constraints on the size of the gamma-ray production
site, supporting a scenario of a very compact emitter. The concept of a blob as a compact emitter inside jets is often used to interpret the
nonthermal emission of AGN. This paradigm is supported by a broad range of observations, in particular bright knots moving
downstream the jets. The origin of such blobs is not clear. It was suggested that the blobs can form when an external
object (e.g. star or a cloud; see \citealt{abr10,bab10} and references therein)
penetrates into the jet and can lose its atmosphere due to the jet impact. Because of
the much larger obstacle  density and the velocity difference, the intruding
matter cannot immediately dilute within the jet but has to be first heated and
accelerated, which in supersonic jets will lead to strong shocks that can
accelerate particles. Therefore, this scenario naturally presents specific
characteristics that are hard to introduce in pure jet models. For instance, the
large blob density makes possible the formation of a  radiation component
typically suppressed in AGN, i.e. emission generated at proton-proton
collisions \citep[suggested to be behind the TeV emission detected from M87 by
HESS][]{bba12}.  Moreover, since the size of the blobs produced in this
scenario is not constrained by the gravitational radius of the central black
hole, it was proposed that this process could be responsible for the very
rapidly variable TeV emission detected from PKS~2155-304 \citep{babkk10,ah07pks}.
The JRGI scenario may have quite general applications, but the properties of the
emission expected within its framework are quite strictly constrained. In particular, the
available energy budget and expected lightcurve shape can be determined with a
simple dynamical model, which describes the evolution of the blob in the jet.
Depending on the jet properties and other factors like the viewing
angle, the dominant emission channel can vary. However, the majority of
the physical conclusions that can be inferred in the JRGI scenario are quite general
because they concern the dynamics of compact blobs in relativistic jets, independently of
the origin of the former and the nature of the radiation channel.

The observations of \c454 with {\it Fermi} revealed several quite
puzzling features, in particular the peculiar lightcurve, with a
nearly steady plateau phase that was interrupted by an exceptionally
bright flare. The detected flux corresponds to an apparent luminosity
of $2\times10^{50}\rm \, erg s^{-1}$, which almost unavoidably implies
a presence of a very powerful jet \citep[see e.g.][]{bgf11}. In the
case of powerful jets, the JRGI scenario should proceed in a quite
specific way as compared to other cases already considered in the
literature \citep{babkk10}. In particular, the mass of the material
initially removed from the star might be very large, resulting in
rather long cloud expansion and acceleration timescales, the main
flare being significantly delayed with respect to the moment of the
star entrance into the jet. The plateau emission would otherwise start
just after the jet penetration, and come from the jet crushing of
lighter clouds ejected from the stellar surface while the star travels
through the jet. The duration of the plateau phase would be determined
by the time required by the main cloud to expand and accelerate.  

We have studied the lightcurve obtained with {\it Fermi} in the context
of the JRGI scenario aiming to satisfy three main properties of the flare: total
energy, maximum luminosity and duration of the plateau stage. It was shown that
the key properties of the jet, i.e. the jet ram pressure (linked to its luminosity) and Lorentz
factor, as well as the cloud/blob characteristics, i.e. mass and cross section, can be
reconstructed as functions of the dimensionless parameter $D$. It
was also shown that in the limit of small $D$-values, the parameter space
is less demanding concerning the jet luminosity, and the key characteristics of the model
saturate at values independent of $D$, which allows conclusive cross-checks of
the scenario. In particular, the flare raise time appeared to be an independent parameter, with its value of $5\,$h closely matching the
rising time of of $4.5\,$h obtained observationally.  Furthermore, it was shown (see Sec.~\ref{pla}) that for the inferred
jet properties the jet-induced stellar wind can provide a mass-loss rate large enough to
generate a steady emission component with a luminosity comparable to that of the
plateau.

Although the analysis of different radiation channels involves
additional assumptions regarding the spectrum of the nonthermal
particles and density of the target fields, it was possible to show
that for magnetic fields not far below equipartition (as expected in a
magnetically launched jet; see Appendix for details) all the conventional radiation channels can be discarded, and  the emission
detected with {\it Fermi} can be produced through proton synchrotron
emission (unless $\eta \rightarrow 1$, making electron synchrotron also feasible). We note that in this case the emission
from pairs created within the blob may also explain the reported optical enhancement at the flare epoch.  

If the
studied scenario is behind the November 2010 flare from \c454, it has
some important implications on the properties of AGN jets. On one
hand, the obtained solution implies a super
Eddington jet. Although this requirement may appear somewhat extreme,
given the exceptional properties of the source it cannot be ruled out
(and in fact super Eddington jets might be a rather common phenomenon;
\citealt{lp12}). In addition, the obtained solution (Eqs.~(\ref{eq:lum_final})~--~(\ref{eq:size_final}))
implies a lower limit on the central black hole mass that, for a
location of the interaction region at $z_0\sim10^{17}\rm cm$, is
roughly consistent with the measured values. Also, adopting a
magnetically launched jet model, it is possible to infer the mass of
the central black hole as $M_{\rm BH,9}=0.5 z_{0,17}^{1/2} \Gamma_{\rm j,1.5}^{-1}$, which is again consistent with the measurements.

At the initial stage of the cloud acceleration, the intensity of the jet-cloud
interaction is the highest, but the associated emission is not Doppler boosted
and therefore very hard to detect in the high-energy regime. Nevertheless, there
are other manifestations  of the JRGI scenario that may be observed.  In
particular, as reported by \citet{lcp13}, a significant enhancement of the line emission
from the source was recorded during the flare epoch. The detected flux was
$\sim2\times10^{45}\rm\, erg\, s^{-1}$, implying an ionizing luminosity at the
level of $10^{46}\rm\, erg\, s^{-1}$. According to our estimates, the initial
blob acceleration period should be characterized by luminosities of $\sim
2\times10^{48}\rm\, erg\, s^{-1}$, so only 1\% of this luminosity would be
required to explain the ionization line component. With the acceleration of the
blob, its radiation gets beamed along the jet, thus reducing its impact on the
BLR and thereby weakening any related line emission.

Since the duration of the expansion phase determines the delay
between the onset of the plateau phase and the flare itself, it is
important to check whether the suggested scenario is consistent with
other flares registered with {\it Fermi} from the source, e.g., in
December 2009 and April 2010 \citep{agile10,fermi10}.  This issue can
be addressed through a simple scaling that relates the duration of the
plateau phase to the total energy released during the active phase:
$t_{\rm pl}\propto E_{\rm tot}^{1/2}$ (assuming a steady jet one can obtain from
equations \eqref{eq:ap_ram} and \eqref{eq:ap_gamma_final} that
$P_0\times\Gamma_0^3=\rm const$, which yields in the scaling dependence). Therefore, for
the previous events, with energy releases 1-2 orders of magnitude smaller
than that of the November 2010 flare,  a
rough estimation of the plateau duration 
gives plateau durations between $1.3$ and $4$ days, consistent with observations.

The amount of RGs or young protostars in the vicinity of the central BH may be enough to 
expect about one JRGI event per year \citep[see estimates in][and references therein]{bab10,babkk10}. However, these estimates 
are unfortunately strongly dependent on the stellar density in the region, a quantity still highly uncertain. Nevertheless, our 
study (and before \citealt{babkk10}) shows that the penetration of a star into the base of the jet in a powerful blazar leads 
to a very characteristic dynamical evolution. In fact JRGI has a quite constrained physics with basically one free parameter: 
the nonthermal efficiency. The specific conditions in the jet/blob interaction region will determine, as long as particle 
acceleration occurs, the dominant radiation process, but in any case powerful high-emission seems natural in the proposed 
scenario.

\appendix

\section{The Structure of Magnetically Driven Jets}\label{acac}

Although the process of jet formation is not fully understood, recent hydrodynamical studies by different groups have shown
that the BZ process may be at work in AGN, and suggest that, once formed, jets may be magnetically accelerated. In that
scenario, the jet base is expected to be magnetically dominated at $z\le 1\mbox{ pc}$ \citep{kbvk07,bk08,bes10}.

During the jet propagation, the magnetic field energy can be transformed into bulk kinetic energy, with a simple
prescription for the bulk Lorentz factor at the linear acceleration stage of the form \citep{bn06} 
\begin{equation}
\Gamma_{\rm j}\approx \frac{\omega}{4 r_{\rm g}}\,,
\label{gam_m}
\end{equation}
where $\omega$ and $r_{\rm g}$ are the jet and the BH gravitational radius, respectively. The opening angle of the
jet is expected to be $\theta=\omega/z\approx 1/\Gamma_{\rm j}$ \citep{kvkb09}. This leads to few useful relations between
different jet parameters:
\begin{equation}
\Gamma_{\rm j}^2\approx \frac{z}{4 r_{\rm g}}\,,\quad
\theta^2 z\approx \frac{\omega^2}z\approx\frac z{\Gamma_{\rm j}^2}\approx {4 r_{\rm g}}\,.
\label{eq:ap_mdj}
\end{equation}
It is also useful to express the above relation in the units used all through the paper:
\begin{equation}
\Gamma_{\rm j,1.5}^2=0.17(=0.4^2)z_{0,17}M_{\rm BH,9}^{-1}\,,
\label{eq:ap_mdj1}
\end{equation}
\begin{equation}
\theta_{-1}^2z_{0,17}=0.6M_{\rm BH,9}\,.
\label{eq:ap_mdj2}
\end{equation}
The jet magnetic field in the comoving frame can also be derived:
\begin{equation}
B_{\rm c}\approx \frac{2}{z}\left(\frac{L_{\rm j}}{c}\right)^{1/2}\approx 120 z_{0,17}^{-1} L_{j,48}^{1/2}\,\mbox{G}\,.
\label{bjc}
\end{equation}

\section{On blob luminosity in the case of heavy blobs}\label{ap:energy}
If the mass of the blob is high, implying a low value of the $D$ parameter, then blobs can travel a significant distance along 
the jet. Therefore, the jet properties can change, and this should have an impact on the emission associated with the blobs. 
These effects can be estimated via the approach suggested by \citet{babkk10}. The dynamics of the blob is characterized by
\begin{equation}
{d\Gamma_{\rm b}\over dt}=\left(\Gamma_{\rm b}^{-2}-\Gamma_{\rm b}^2\Gamma_{\rm j}^{-4}\right){P_{\rm j} \pi r_{\rm b}^2\over 
4cM_{\rm b}}\,,
\end{equation}
where $P_{\rm j}$ is the ram pressure of the jet.
Since the emission of the blob is important only when it moves relativistically, it is safe to take $z=z_{\rm 0}+ct$ and the 
initial condition $\Gamma_{\rm b}|_{t=0}=1$. The above equation can be modified to a dimensionless form: 
\begin{equation}
{dg\over dy}=\left({\left(\Gamma_{\rm j}/\Gamma_0\right)^2\over g^2}-{g^2\over \left(\Gamma_j/\Gamma_0\right)^2}\right){D\over 
\left(\Gamma_j/\Gamma_0\right)^2\left(P_{\rm 0}/P_{\rm j}\right)}\,,
\label{eq:motion_eq}
\end{equation}
where $z=z_0y$; $\omega_0$, $\Gamma_0$ and $P_{\rm 0}$ are the jet radius, ram pressure and Lorentz factor at $z=z_0$; 
$g=\Gamma_{\rm b}/\Gamma_0$; and 
the dimensionless parameter $D$ is determined as
\begin{equation}
D={P_{\rm 0}\pi r_{\rm b}^2z_0\over4c^2M_{\rm b}\Gamma_0^3}\,.
\label{eq:D_app}
\end{equation}
This approach is nearly identical to the one developed earlier, but here one accounts for the possible change of the properties 
of the jet (e.g. for the increase of the jet Lorentz factor). 

If $D\gg 1$, the blob gets rapidly involved into the jet motion, so the solution is identical to the one obtained in  
\citet{babkk10} even if the properties of the jet are changing with z. However, if $D<1$, the structure of the jet may have 
some influence on the properties of the emission. In particular, for purely magnetically driven jets, i.e. with a parabolic 
shape, the above equation can be simplified, since the following relations hold
$$
\omega_0=2\sqrt{r_{\rm g}z_0}\,,
$$
$$
\omega=\omega_0 y^{1/2}\,,\quad P_{\rm j}=P_{\rm 0} y^{-1}\,,
$$
$$
\Gamma_{\rm j}=z_0/\omega_0 y^{1/2}=\Gamma_0y^{1/2}\,.
$$
Thus, the equation of motion (\ref{eq:motion_eq}) yields
\begin{equation}
{dg\over dy}=\left({y\over g^2}-{g^2\over y}\right){D\over y^2}\,,
\end{equation}
with the boundary condition $g|_{y=1}=\omega_0/z_0$.  Although this equation does not have any analytical solution, an {\it 
asymptotic} solution can be derived for $D\ll1$ (i.e. the jet structure is important):
\begin{equation}
g_{\rm ap}=\left[\left(\omega_0/z_0\right)^3+3D\ln{y}\right]^{1/3}\,.
\end{equation}

For the case of a conical jet (i.e., $\Gamma_{\rm j}={\rm const}$, $\omega=z/\Gamma_{\rm j}$ and $P_{\rm j}=P_{\rm 0}y^{-2}$), 
the equation of motion is reduced to the equation:
\begin{equation}
{dg\over dy}=\left({g^{-2}}-{g^2}\right){D\over y^2}\,,
\end{equation}
which allows an analytical solution \citep[for details, see][]{babkk10}. However, since the analytical solution is rather 
bulky,  we use the asymptotic solution:
\begin{equation}
g_{\rm ac}=\left[\left(\omega_0/z_0\right)^3+3D(y-1)/y\right]^{1/3}\,.
\end{equation}
which is valid for $D\ll1$. Since the models of parabolic and conical jets correspond to the most feasible jet model, we will 
consider the realization of the JRGI scenario for these two cases. The obtained asymptotic solutions allow one to clearly estimate 
the differences that depend on the specific jet configuration. 

The intensity of the nonthermal emission of the blob, corrected for Doppler boosting, has the following form:
\begin{equation}
L_\gamma=4L_{\rm j}\left({r_{\rm b}\over \omega_0}\right)^2\Gamma_0^2\,F_{\rm e}\left(y\right)\,,
\label{eq:ap_lum}
\end{equation}
where $F_{\rm e}\left(y\right)=\,{g^4\over \left(P_{\rm 0}/P_{\rm j}\right)}\left({1\over g^2}-{g^2\over \left(\Gamma_{\rm 
j}/\Gamma_0\right)^4}\right)$ is the correction function.

The maximum value of  the correction function, $F_{\rm e,max}$, depends on the value of the $D$ parameter and on the structure 
of the jet. If $D\geq1$  it saturates on a constant value of $F_{\rm e,max}\approx0.4$ independent of the jet properties. In 
the case of $D\ll1$,  the asymptotic  solutions show that the maximum of the correction function is reached at $y_{\rm 
m}\approx1.95$, with $F_{\rm e,max}\approx0.82D^{2/3}$ for a parabolic jet, and for a conical jet, $y_{\rm m}\approx1.33$ and 
$F_{\rm e,max}\approx0.46D^{2/3}$. In the intermediate range, $0.1<D<1$, $F_{\rm e,max}$ is $\sim 0.5D^{0.5}$ and $\sim 
0.3D^{0.5}$ for the parabolic and conical jets, respectively. The location of the maximum $y_{\rm m}$ has a weak dependence of 
$D$ in this regime; i.e. depending on the structure of the jet, the maximum intensity, $F_{\rm e,max}$, can vary by a factor of 
$\sim 2$ in the case of $D<1$.

The total energy emitted by a blob can be obtained via the integration of the luminosity over the observer time 
($d\tau=z_0dy/(2c\Gamma_{\rm b}^2)$):
\begin{equation}
E_{\gamma}\simeq 8\xi\left(\int\limits_1^\infty dy F_{\rm e}{D\over g^2}\right)M_{\rm b/c}c^2\Gamma_{\rm j}^3=8\xi \bar{F}_{\rm 
e}M_{\rm b/c}c^2\Gamma_{\rm j}^3\,.
\label{eq:toten_ap}
\end{equation}
This equation {\it defines} the function $\bar{F}_{\rm e}$.

For large values of the $D$ parameter, the integration term in the above equation can be obtained analytically if one accounts 
for the
relation $F_{\rm e}=(dg/dy)g^4/D$:
\begin{equation}
\bar{F}_{\rm e}=\int\limits_1^\infty dy F_{\rm e}{D\over g^2}=\int\limits_0^{g_{\rm max}} dg\, g^2=\frac13g_{\rm max}^3\,.
\end{equation}
Here we note that, in the case of an accelerating jet, the integral determining the total emitted energy is divergent. 
Formally, this is so because of $g_{\rm max}\rightarrow\infty$. Therefore, the integration should be artificially truncated at 
some point. Physically, this point would correspond to the moment when the blob gets homogenized in the jet. For large values 
of $D$, the blob speed rapidly reaches the jet velocity, and one should adopt $g_{\rm max}=1$. 
For small $D$ values, we use $\int\limits_1^\infty dy F_{\rm e}{D\over g^2}=D$, which is a precise identity for the case of a 
conical jet, and corresponds to a truncation at $y\sim 3$ for a parabolic jet. Therefore, for the present order-of-magnitude 
estimates, one can estimate the energy emitted by the blob as
\begin{equation}
\bar{F_{\rm e}}=\int\limits_1^\infty dy F_{\rm e}{D\over g^2}=\min\left(D,\frac13\right)\,.
\end{equation}

Since the structure of the jet only imposes uncertainties of the order of 2, for the order-of-magnitude estimates presented in 
this paper we use the solution obtained for the conical jet structure. To use the solution obtained for the parabolic structure 
of the jet one should properly describe the process of the blob homogenization in the jet, which is out of the scope of this 
work. 


\begin{figure*}
\vspace{0.6cm}
\includegraphics[width=0.99\textwidth,angle=0]{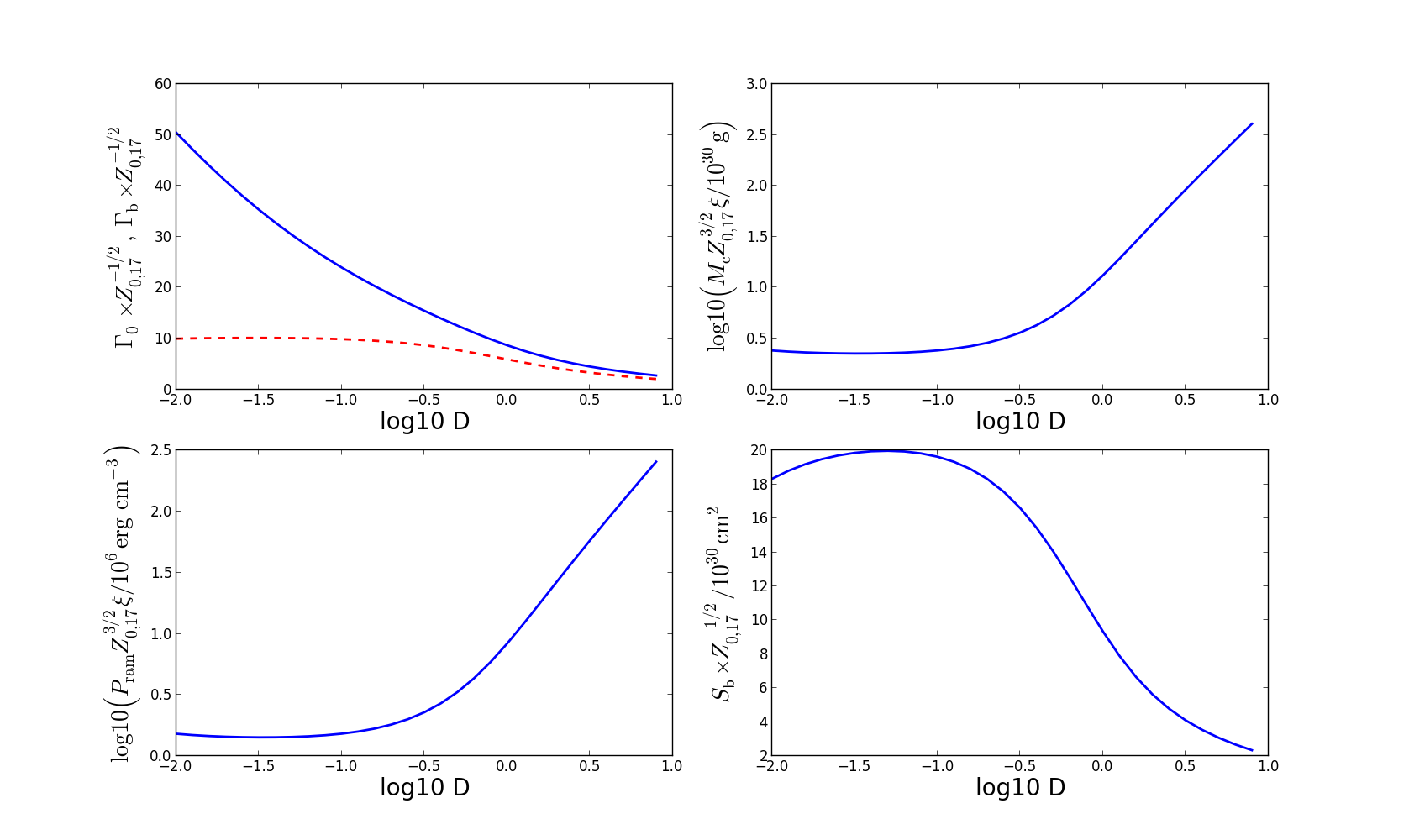}
\vspace{0.5cm}
\caption{The derived jet parameters as a function of the $D$ parameter.}
\label{fig:solution}
\end{figure*}

\section{A solution for the Jet Parameters}\label{ap:solution}

One can obtain a simplified analytical solution which allows the determination of 
the jet parameters. 
Ignoring the weak dependence on $R_*$ in equation~\eqref{eq:ex_time_2}, one obtains:
\begin{equation}
M_{\rm c,30}={3\times10^{-2}\over \xi F_{\rm e,max}\Gamma_{\rm j,1.5}^{2}S_{\rm b,32}}
\left({L_\gamma\over 2\times10^{50}\rm erg}\right)\left({t_{\rm pl}\over 13\rm\, d}\right)^{-2}\left({1+z_{\rm 
rs}\over1.859}\right)^2\,.
\label{eq:ap_flux}
\end{equation}
From equation~(\ref{mass_condtion}), one can infer the mass of the cloud: 
\begin{equation}
M_{\rm c,30}={0.1\over\xi \bar{F}_{\rm e}\Gamma_{\rm j,1.5}^3}\left({E_\gamma\over 2\times10^{55}\rm erg}\right)\,,
\label{eq:ap_mass}
\end{equation}
and equation~(\ref{eq:super_edd}) can be re-normalized as
\begin{equation}
P_0=2\times10^4  \xi^{-1} F_{\rm e,max}^{-1}\Gamma_{0,1.5}^{-2}S_{\rm b,32}^{-1}\left({L_\gamma\over 2\times10^{50}\rm
erg}\right)\rm \,erg\,cm^{-3}\,.
\label{eq:ap_ram}
\end{equation}
Finally, the definition of the $D$ parameter, equation~(\ref{DD}), provides the fourth required equation to determine the 
parameters $P_0$, $S_{\rm b}$, $M_{\rm c}$ and $\Gamma_0$. Here we  assume that the flare was the result of the superposition 
of 4 subflares (i.e., $M_{\rm c}=4M_{\rm b}$):
\begin{equation}
D={6\times10^{-3}z_{0,17}\over F_{\rm e,max}\xi M_{\rm c,30}\Gamma_{0,1.5}^5}\left({L_\gamma\over 2\times10^{50}\rm 
erg}\right)\,.
\label{eq:ap_DD}
\end{equation}
Equations~(\ref{eq:ap_flux}~--~\ref{eq:ap_DD}) have a unique solution, which characterizes the parameters $P_0$, $S_{\rm b}$, 
$M_{\rm c}$ and $\Gamma_0$ as functions of $D$:

\begin{equation}
P_0=3\times10^6\frac{F_{\rm e,max}^{1.5}D^{1.5}}{\bar{F}_{\rm e}^{2.5}\xi z_{0,17}^{1.5}}\left({L_\gamma\over 2\times10^{50}\rm 
erg}\right)^{-1.5}\left({t_{\rm pl}\over 13\rm\, d}\right)^{-2}\left({E_\gamma\over 2\times10^{55}\rm erg}\right)^{2.5}\,\rm 
erg\,cm^{-3}\,
\label{eq:ap_lum_final}
\end{equation}
\begin{equation}
M_{\rm c}=4M_{\rm b}=5\times10^{30}\frac{F_{\rm e,max}^{1.5}{D}^{1.5}}{\bar{F}_{\rm e}^{2.5}\xi 
z_{0,17}^{1.5}}\left({L_\gamma\over 2\times10^{50}\rm erg}\right)^{-1.5}\left({E_\gamma\over 2\times10^{55}\rm 
erg}\right)^{2.5}\,\rm g,
\label{eq:ap_mass_final}
\end{equation}
\begin{equation}
\Gamma_0=8\left( \frac{\bar{F}_{\rm e}z_{0,17}}{F_{\rm e,max}D}\right)^{0.5}\left({L_\gamma\over 2\times10^{50}\rm 
erg}\right)^{0.5}\left({E_\gamma\over 2\times10^{55}\rm erg}\right)^{-0.5}\,,
\label{eq:ap_gamma_final}
\end{equation}
and
\begin{equation}
S_{\rm b}=8\times10^{30}\frac{z_{0,17}^{0.5}\bar{F}_{\rm e}^{1.5}}{F_{\rm e,max}^{1.5}D^{0.5}}\left({L_\gamma\over 
2\times10^{50}\rm erg}\right)^{1.5}\left({t_{\rm pl}\over 13\rm\, d}\right)^2\left({E_\gamma\over 2\times10^{55}\rm 
erg}\right)^{-1.5}\,\rm cm^2.
\label{eq:ap_size_final}
\end{equation}

The values $F_{\rm e,max}$ and $\bar{F}_{\rm e}$ depend on the value of the $D$ parameter and the structure of the jet. To 
illustrate the feasibility of the derived solution, in figure~\ref{fig:solution} we show the jet parameters as functions of the 
$D$ parameter.


\section*{Acknowledgments}
The authors are thankful to S. Kelner for useful discussions and Benoit Lott for providing observational data.  
The research leading to these results has received funding from the European
Union Seventh Framework Program (FP7/2007-2013) under grant agreement
PIEF-GA-2009-252463. 
V.B.-R. acknowledges support by the Spanish 
Ministerio de Ciencia e Innovaci\'on
(MICINN) under grants AYA2010-21782-C03-01 and FPA2010-22056-C06-02. 
V.B.-R. acknowledges financial support
from MINECO through a Ram\'on y Cajal fellowship.
This research has been supported by the Marie Curie Career Integration
Grant 321520.
BMV acknowledges partial  support  by  RFBR  grant  12-02-01336-a.


\end{document}